\documentclass[a4paper,12pt]{article}
\usepackage{cite}
\usepackage{amssymb}
\newcommand {\oks}[2]{{\raise0.7ex\hbox{${\scriptstyle #1}$}\!\mathord{\left/
{\vphantom{{1}{2}}}\right.\kern-\nulldelimiterspace}\!\lower0.7ex
\hbox{${\scriptstyle #2}$}}}
\begin{document}

\title{New effects in neutrino oscillations in matter \\ and
electromagnetic fields\thanks{A short version of invited talk
given at the 4th International Bruno Pontecorvo School on
"Neutrino Oscillations, CP and CPT Violation: the Three Windows to
Physics Beyond the Standard Model", Capri, May 26-29, 2003.}}

\author{A.Studenikin\thanks{E-mail: studenik@srd.sinp.msu.ru}}
\date{}
\maketitle

\begin{center}
{\em
Department of Theoretical Physics, Moscow State University,
119992, Moscow, Russia}
\end{center}

At  present neutrino oscillation phenomenon is the most evidently
confirmed window to physics beyond the Standard Model. The
thorough studies of neutrino oscillations were launched by the
pioneering paper of Bruno Pontecorvo \cite{Pon57}. Indeed, it is
surprising that after more than 45 years of researches in this
field it is still possible to discover new effects in  neutrino
oscillations.

It is a pleasure for me to present at the 4th International School
dedicated to Bruno Pontecorvo the four new effects in neutrino
oscillations that have been recently studied in my research group
at the Moscow State University. The whole story began several
years ago  when we studied neutrino spin oscillations in strong
magnetic fields \cite{LikStu_JETP95} and to our much surprise
realized that at that time in literature there were no even
attempts to consider neutrino spin oscillations in any
electromagnetic field configurations rather than constant in time
and transversal in respect to neutrino motion magnetic fields
(see, for example, \cite{VolVysOku86, Akh88, LimMar88}). Further
more, in all of the studies of neutrino spin and also flavour
oscillations in matter, performed before 1995, the matter effect
was treated only in the non-relativistic limit.

The first our attempt \cite {LikStu95} to consider neutrino
flavour oscillations in matter in the case when matter is moving
with relativistic speed was made in 1995. In that our study we
tried to apply the Lorentz invariant formalism to describe
neutrino flavour oscillations and realized that the value of the
matter term in the neutrino effective potential can be
significantly changed if matter is moving with relativistic speed.
We have continued \cite{EgoLobStu00} our studies on evaluation of
the Lorentz invariant formalism in neutrino spin oscillations in
1999 and have developed an approach that enables us to consider
neutrino spin oscillations in arbitrary electromagnetic field
configurations. We have shown \cite{EgoLobStu00} how the
Bargmann-Michel-Telegdi equation \cite{BMT59}, describing a
neutral particle spin evolution under the influence of an
electromagnetic field, can be generalized for the case of neutrino
moving in electromagnetic field and matter. The corresponding
Lorentz invariant neutrino spin evolution equation have been
derived from the Bargmann-Michel-Telegdi equation by the following
substitution: one has to add to the electromagnetic field tensor
$F_{\mu \nu}$ an anti-symmetric tensor $G_{\mu \nu}$ that can be
constarcted with use of the neutrino speed, matter speed, and
matter polarization four-vectors under the natural assumptions
that the equation have to be linear over $F_{\mu \nu}$ and over
the other mentioned above vectors.

From this new generalized BMT equation for the neutrino spin
evolution in an electromagnetic field and matter the corresponding
effective Hamiltonian describing neutrino spin oscillations is
just straightforward \cite{EgoLobStu00}. Thus, the first new
effect is the prediction of neutrino spin oscillations in various
electromagnetic field configurations. We have derived new
resonances in neutrino oscillations for several electromagnetic
fields such as the field of an electromagnetic wave and
superposition of an electromagnetic wave and constant magnetic
field.

We have predicted \cite{EgoLobStu00,LobStu01}, using the
generalized BMT equation, the second new effect: neutrino spin
procession can be stimulated not only by presence of
electromagnetic fields (i.e., by electromagnetic interactions of
neutrino) but also by weak interactions of neutrino with
background matter. Moreover, we have shown \cite{LobStuplb_03}
that the neutrino spin precession always occurs in presence of
matter (even in the case of non-moving and unpolarized matter) if
the initial neutrino state is not longitudinally polarized. We
have also considered \cite{DvoStu02} the neutrino spin evolution
problem within the Lorentz invariant approach in the case of
general types of neutrino non-derivative interactions with
external fields.

Then we have used the derived new equation for neutrino spin
evolution for the case when matter may move with arbitrary (also
relativistic) speed and may be polarized. We have predicted
\cite{LobStu01} the third effect: the matter motion can
drastically change the neutrino oscillation pattern and, in
particular, can sufficiently shift the neutrino spin oscillation
resonance condition \cite{Akh88,LimMar88}, if compared with the
case of non-moving matter. The analogous effect exist
\cite{LikStu95,GriLobStuplb_02} also in neutrino flavour
oscillations: the neutrino resonance condition \cite{Wol78,
MikSmi85} can be significantly modified if matter is moving with
relativistic speed.

The fourth new effect in neutrino oscillations is the prediction
\cite{LobStuplb_03} for the new mechanism of electromagnetic
radiation by neutrino moving in background matter and/or
electromagnetic fields. This new mechanism of electromagnetic
radiation originates from the neutrino spin precession that can be
produced whether by weak interactions with matter or by
electromagnetic interactions with external electromagnetic fields.
We have named this radiation as "spin light of neutrino"
($SL\nu$). The total power of the $SL\nu$ does not washed out when
the emitted photon refractive index is equal to unit and the
$SL\nu$ can not be considered as the neutrino Cerenkov radiation
(see, for example, \cite{IoaRaf97} and references therein). The
discovered important properties of $SL\nu$ (such as strong beaming
of the radiation along the neutrino momentum, the rapid growth of
the total radiation power with the neutrino energy and density of
matter, the possibility to emit photons with energies span up to
gamma-rays) enables us to predict that this radiation should be
important in different astrophysical environments (quasars,
gamma-ray bursts etc) and in dense plasma of the early Universe.

\end{document}